\documentclass[sn-basic]{sn-jnl}
\usepackage[fontsize=11pt]{fontsize}

\usepackage{graphicx}%
\usepackage{multirow}%
\usepackage{amsmath,amssymb,amsfonts}%
\usepackage{amsthm}%
\usepackage{mathrsfs}%
\usepackage[title]{appendix}%
\usepackage{xcolor}%
\usepackage{textcomp}%
\usepackage{manyfoot}%
\usepackage{booktabs}%
\usepackage{algorithm}%
\usepackage{algorithmicx}%
\usepackage{algpseudocode}%
\usepackage{listings}%
\usepackage{verbatim}
\usepackage{xcolor}

\theoremstyle{thmstyleone}%
\theoremstyle{thmstyletwo}%

\theoremstyle{thmstylethree}%

\begin{document}

\title[Limitations of Validity Intervals in Data Freshness Management]{Limitations of Validity Intervals in Data Freshness Management}


\author{\fnm{Kyoung-Don}
\sur{Kang}}\email{kang@binghamton.edu}



\affil{\orgdiv{Department of Computer Science},  
\orgname{State University of New York at Binghamton}}




\abstract{In data-intensive real-time applications, such as smart transportation and manufacturing, ensuring data freshness is essential, as using obsolete data can lead to negative outcomes. Validity intervals serve as the standard means to specify freshness requirements in real-time databases. In this paper, we bring attention to significant drawbacks of validity intervals that have largely been unnoticed and introduce a new definition of data freshness, while discussing future research directions to address these limitations.}





\maketitle

\section{Introduction}\label{sec:intro}
In data-intensive real-time applications, such as smart transportation, manufacturing, and healthcare, real-time data analysis using fresh data that reflects the current real-world conditions is essential. 
The notion of data freshness, namely, the data temporal consistency, was introduced in real-time databases (RTDBs) for temporal data, exemplified by sensor data, which have the potential to change over time \cite{ramam93}. 
At time $t$, a temporal data item $O_i$ in the system is considered fresh if $t \le u(O_i) + VI(O_i)$, where $u(O_i)$ is the most recent update time of $O_i$ and $VI(O_i)$ is its validity interval. 
Presently, this approach stand as the established standard for enforcing data freshness requirements in RTDBs.

\textbf{Striving to meet data freshness constraints based on validity intervals, however, may prove infeasible and lead to potential failures.} Consider a real-time transaction, $\tau_i$, responsible for retrieving and analyzing an arbitrary temporal data element $O_i$. 
If $VI(O_i)$ expires before the analysis concludes, $\tau_i$ will reacquire and reanalyze $O_i$ \cite{kdtkde04, kim-rtsj19}.
In an extreme situation, this cycle can continue indefinitely and miss the deadline of $\tau_i$, if the data retrieval time $R(O_i)$, analysis time $A(O_i)$, or $R(O_i) + A(O_i)$ in $\tau_i$ exceeds $VI(O_i)$. 
In particular, this may occur even when $\tau_i$ is the only transaction in the system, because supporting data freshness via validity intervals is not associated with any explicit feasibility check to avoid such a devastating situation.
The problem is aggravated when many transactions contend for data and resources.

\textit{The problem becomes serious when dealing with computationally intensive transactions for intelligent real-time data services}, such as object detection and tracking via deep learning, or data retrieval over unreliable wireless connections in CPS or IoT. 
Despite its significance, scant attention has been paid to this problem. 

Furthermore, temporal data in RTDBs are updated periodically to uphold their freshness as per their validity intervals \cite{ramam93, rtdbBook, li-tc16}, \textit{regardless of actual  data access patterns and data value changes}. Thus, temporal data are updated periodically even when they are not accessed at all or their value changes are negligible. 


This paper outlines research directions to address these challenges in supporting sensor data freshness and the timeliness of intelligent real-time sensor data analysis in CPS and IoT.

\section{Research Directions for Effective Freshness Management}
\label{sec:fvi}

In this section, we introduce a feasibility condition to meet data freshness constraints in data-intensive CPS and IoT applications, and sketch related research directions.





\subsection{Feasibility Condition for Data Freshness Guarantee}

If an arbitrary transaction $\tau_i$ is responsible for analyzing a set of temporal data $S_i$, we introduce \textbf{a new feasibility condition for data freshness} as $VI(O_i) \ge R_i(O_i) + A(O_i), \forall O_i \in S_i$. This essential check is necessary to prevent the indefinite restarting of $\tau_i$ caused by the premature expiration of $VI(O_i)$. 
Unlike the classical definition of validity intervals \cite{ramam93}, this definition requires that the validity interval of $O_i$ be not shorter than the sum of the data retrieval and analysis time.
Historically, this feasibility condition has been implicitly met, primarily because data retrieval and analysis times were negligible and predictable within centralized real-time embedded systems. Examples of such systems include industrial control loops and electronic control units in a vehicle that compare readings from local sensors (e.g., temperature and pressure sensors) against predefined thresholds and generate control signals to maintain system stability.

However, \textit{satisfying the feasibility condition for data freshness presents remarkable challenges} in emerging intelligent real-time data services, such as smart transportation and manufacturing: 
\begin{itemize}
    \item A sophisticated analysis of big temporal data, such as analyzing a real-time video or images via machine learning, may take a long time in resource-constrained embedded systems. As a result, $VI(O_i)$ can be exceeded before the analysis of $O_i$ completes.

    \item The time required for data analysis can significantly fluctuate based on data values, primarily due to conditional data processing and associated branches. Even when the maximum $A(O_i)$ is predetermined (e.g., on the critical path) for a transaction $\tau_i$ when executed in isolation, it can be aborted/restarted at runtime due to data/resource contention with other transactions. As a result, $VI(O_i)$ may expire before $\tau_i$ commits, causing $\tau_i$ to restart.
    

    \item Predicting the time for data retrieval is challenging, especially when sensor data is gathered via unreliable wireless media.
    
\end{itemize}

Despite a significant body of work on real-time scheduling \cite{jliu00,but11,sanjoy14,sanjoy-dt18}, \textit{little prior work has been done to meet the feasibility condition for data freshness proposed in this paper} ($VI(O_i) \ge R_i(O_i) + A(O_i), \forall O_i$). 
Thus, research on novel scheduling methodologies to ensure the freshness of temporal data while meeting transaction deadlines at the same time is required. Another important research direction in this context is to explore how to reduce data conflicts and update workloads in RTDBs. Several approaches are outlined in the following subsections.



\subsection{Multi-Version Approach to Temporal Data} 
The definition of validity intervals enables real-time data analysis transactions to consider only the current snapshot of the real-world state, neglecting a continuous time series depicting its evolution. This limitation poses a risk of failure in real-time data analysis, where transactions may repeatedly restart due to the expiration of validity intervals before completion. 

To tackle this challenge, we advocate for the adoption of multiple versions for an arbitrary temporal data object $O_i$, \textit{ensuring temporal consistency based on $VI(O_i)$ concerning the time when a specific transaction $\tau_i$ accesses $O_i$}. Assume that $O_i$ is updated at time $t_1$ and accessed by $\tau_i$ at a later time $t_2$, where $t_1 \le t_2 \le t_1 + VI(O_i)$. If $VI(O_i)$ expires at $t_3 > t_2$ before $\tau_i$ commits, a new version of $O_i$ is created upon update. However, $\tau_i$ is allowed to continue data analysis without restarting, as $O_i$ was considered fresh concerning $\tau_i$ at the time of its access, i.e., $t_2$. 
    

Multi-version concurrency control (MVCC) protocols have been explored in RTDBs \cite{tom05,kim1991enhancing}; however, the problem of potential failures incurred by validity interval expirations has not been delved into. Therefore, real-time MVCC protocols could be extended to avoid validity interval-induced (VI-induced) failures, while supporting the serializability of concurrent transactions. 
    
    

\subsection{Methods for Decreasing Temporal Data Updates} 

An inherent drawback of the multi-version model is the increased resource usage required to manage at least two versions of a temporal data item: the version that was fresh when one or more user transactions accessed it and the new version produced as a result of an update. A possible approach integrating the multi-version approach that aims to prevent VI-induced failures with effective methods explored independently to reduce update workloads without intending to handle VI-caused failures:
    \begin{itemize}
        \item \textit{On-Demand Updates for Temporal Data}: To ensure data freshness, periodic updates are applied to temporal data, irrespective of whether they are being accessed by any user transaction \cite{ramam93, rtdbBook, li-tc16}. In data-intensive systems, however, user transactions often access a subset of data \cite{kd-rtsj18}.
        For example, traffic data at busy intersections or severe weather data are typically accessed more frequently. 
        Thus, updating temporal data on demand can reduce update workloads while maintaining the freshness of the accessed data \cite{kdeuro02,kim-rtss16,kim-rtsj19}.

        \item \textit{Flexible Validity Intervals}: The concept of flexible validity intervals was initially proposed in \cite{kdtkde04}. It entails extending the validity intervals of cold data, characterized by a higher update frequency than access frequency, by applying the principles of the elastic task model \cite{elastic} with a consideration of data access patterns. 
        Similarly, the $(m,k)$-firm model \cite{mk} can be utilized to reduce the frequency of updating cold data in a controlled manner, while extending the validity intervals of cold data objects accordingly.

        \item \textit{Adaptive Update Methods Leveraging Data Redundancy}: The value of a temporal data object may not significantly vary from one update period to the next; therefore, updating similar data can be skipped to reduce the load \cite{similarity}.
        Moreover, utilizing a data prediction model, as demonstrated in \cite{cheng2019data}, can help minimize bandwidth consumption by transmitting only essential data that are anticipated to considerably deviate from the values predicted by the model shared between the data source and the sink, such as the RTDB. 
        Hence, approaches based on data similarity or prediction complement the definition of validity intervals that does not account for actual data values and their fluctuations. 
    
    \end{itemize}




Past studies have explored strategies for MVCC and diminishing update workloads. However, most existing research has overlooked the possibility of failures in RTDBs due to unfeasible validity interval requirements. Additionally, there is a scarcity of prior efforts aimed at exploring a comprehensive approach to forestall VI-induced failures, while simultaneously minimizing update workloads. How to effectively bridge the gap is a predominantly open problem.

\section{Conclusions and Future Work}\label{sec:conc}

Ensuring the freshness of temporal data is essential in data-intensive real-time applications.
Since its introduction (\cite{ramam93}), the notion of validity intervals have been widely employed to meet data freshness constraints in data-intensive real-time applications. Nevertheless, it has serious limitations that may result in failures and waste of resources precious in real-time embedded systems, when directly applied to emerging intelligent real-time data services such as smart transportation. In this paper, our objective is to bring attention to this critical issue and propose potential directions for future research to tackle it.

\bibliography{02_kd-refs}

\begin{thebibliography}{18}
\providecommand{\natexlab}[1]{#1}
\providecommand{\url}[1]{{#1}}
\providecommand{\urlprefix}{URL }
\providecommand{\doi}[1]{\url{https://doi.org/#1}}
\providecommand{\eprint}[2][]{\url{#2}}
 \bibcommenthead

\bibitem[{Baruah(2018)}]{sanjoy-dt18}
Baruah S (2018) Mixed-criticality scheduling theory: Scope, promise, and limitations. {IEEE} Design {\&} Test 35(2):31--37

\bibitem[{Baruah et~al(2014)Baruah, Bertogna, and Buttazzo}]{sanjoy14}
Baruah S, Bertogna M, Buttazzo G (2014) {Multiprocessor scheduling for real-time systems}. Springer

\bibitem[{Buttazzo(2011)}]{but11}
Buttazzo G (2011) {Hard Real-Time Computing Systems: Predictable Scheduling Algorithms and Applications}, 3rd edn. Springer

\bibitem[{Buttazzo et~al(1998)Buttazzo, Lipari, and Abeni}]{elastic}
Buttazzo G, Lipari G, Abeni L (1998) {Elastic Task Model For Adaptive Rate Control}. In: IEEE Real-Time Systems Symposium

\bibitem[{Cheng et~al(2019)Cheng, Xie, Wu, Yu, and Li}]{cheng2019data}
Cheng H, Xie Z, Wu L, et~al (2019) {Data Prediction Model in Wireless Sensor Networks based on Bidirectional LSTM}. EURASIP Journal on Wireless Communications and Networking 2019:1--12

\bibitem[{Gustafsson et~al(2005)Gustafsson, Hallqvist, and Hansson}]{tom05}
Gustafsson T, Hallqvist H, Hansson J (2005) {A Similarity-Aware Multiversion Concurrency Control and Updating Algorithm for Up-To-Date Snapshots of Data}. In: Euromicro Conference on Real-Time Systems

\bibitem[{Hamdaoui and Ramanathan(1995)}]{mk}
Hamdaoui M, Ramanathan P (1995) {A Dynamic Priority Assignment Technique for Streams with (m,k)-Firm Deadlines}. IEEE Transactions on Computers 44(12)

\bibitem[{Ho et~al(1997)Ho, Kuo, and Mok}]{similarity}
Ho SJ, Kuo TW, Mok AK (1997) {Similarity-based Load Adjustment for Real-Time Data-Intensive Applications}. In: IEEE Real-Time Systems Symposium

\bibitem[{Kang(2018)}]{kd-rtsj18}
Kang KD (2018) {Enhancing Timeliness and Saving Power in Real-Time Databases}. Real-Time Systems 54(2):484--513

\bibitem[{Kang et~al(2002)Kang, Son, Stankovic, and Abdelzaher}]{kdeuro02}
Kang KD, Son SH, Stankovic JA, et~al (2002) {A QoS-Sensitive Approach for Timeliness and Freshness Guarantees in Real-Time Databases}. In: Euromicro Conference on Real-Time Systems

\bibitem[{Kang et~al(2004)Kang, Son, and Stankovic}]{kdtkde04}
Kang KD, Son S, Stankovic JA (2004) {Managing Deadline Miss Ratio and Sensor Data Freshness in Real-Time Databases}. IEEE Transactions on Knowledge and Data Engineering 16(10):1200--1216

\bibitem[{Kim et~al(2016)Kim, Abdelzaher, Sha, Bar-Noy, and Hobbs}]{kim-rtss16}
Kim JE, Abdelzaher TF, Sha L, et~al (2016) {Sporadic Decision-centric Data Scheduling with Normally-off Sensors}. In: IEEE Real-Time Systems Symposium

\bibitem[{Kim et~al(2019)Kim, Abdelzaher, Sha, Bar{-}Noy, Hobbs, and Dron}]{kim-rtsj19}
Kim JE, Abdelzaher TF, Sha L, et~al (2019) {Decision-Driven Scheduling}. Real-Time Systems 55(3):514--551

\bibitem[{Kim and Srivastava(1991)}]{kim1991enhancing}
Kim W, Srivastava J (1991) {Enhancing Real-Time DBMS Performance with Multiversion Data and Priority Based Disk Scheduling}. In: IEEE Real-Time Systems Symposium, pp 222--223

\bibitem[{Lam and Kuo(2006)}]{rtdbBook}
Lam KY, Kuo TW (eds)  (2006) {Real-Time Database Systems}. Kluwer Academic Publishers

\bibitem[{Li et~al(2016)Li, Chen, Xiong, Li, and Wei}]{li-tc16}
Li J, Chen J, Xiong M, et~al (2016) {Temporal Consistency Maintenance Upon Partitioned Multiprocessor Platforms}. IEEE Transactions on Computers 65(5):1632--1645

\bibitem[{Liu(2000)}]{jliu00}
Liu JWS (2000) {Real-Time Systems}. Prentice Hall

\bibitem[{Ramamritham(1993)}]{ramam93}
Ramamritham K (1993) {Real-Time Databases}. International Journal of Distributed and Parallel Databases 1(2)

\end{thebibliography}

\section*{Acknowledgement}
This work was partially supported by National Science Foundation grants CNS-2007854 and CNS-2326796. 

\end{document}